# Efficient cascaded wavelength conversion under two-peak Stark-chirped rapid adiabatic passage via grating structures

Handa Zhang, Xiang Zhang, Ting Wan, Dong Chen, Fujie Li, Zhonghao Zhang, and Changshui Chen

Guangdong Provincial Key Laboratory of Nanophotonic Functional Materials and Devices & Guangzhou Key Laboratory for Special Fiber Photonic Devices, School of Information Optoelectronic Science and Engineering, South China Normal University, Guangzhou, Guangdong 510006, People's Republic of China

Corresponding author: Changshui Chen (e-mail: 20071053@m.scnu.edu.cn).

**ABSTRACT** In this paper, we demonstrate a domain inversion crystal structure to study the cascaded three-wave mixing process in a two peak Stark-chirped rapid adiabatic passage. We have achieved efficient wavelength conversion, which can be performed in intuitive order and counterintuitive order. The requirement of crystal is reduced and the flexibility of structure design is improved. When the conversion wavelength is fixed, increasing the coupling coefficient between the two peaks can reduce the intensity of the intermediate wavelength while maintaining high conversion efficiency. Compared with the cascaded wavelength conversion process based on stimulated Raman rapid adiabatic passage, the two-peak Stark-chirped rapid adiabatic passage has a larger convertible wavelength bandwidth. This scheme provides a theoretical basis for obtaining mid-infrared laser source via flexible crystal structure.

**KEYWORDS** Two-cascaded wavelength conversion; Optical crystal; Stark-chirped rapid adiabatic passage.

## I. INTRODUCTION

Wavelength conversion, via three-wave mixing (TWM) processes in quadratic nonlinear optical crystals, is widely used to generate laser wavelengths that are not available by direct laser action[1-3]. During a TWM process, the efficiency mainly depends on the fulfillment of the phase-matching condition. Quasi-phase-matching (QPM) is a common method to facilitates control over phase-matching conditions [4, 5]. The sign of the nonlinear coefficient is modulated in the process. The conventional TWM process generally has the disadvantage of low conversion efficiency. Thus, efficient wavelength conversion is expected.

  Suchowski *et al*. introduced an adiabatic conversion scheme from a two-level system to achieve high efficiency and wide bandwidth conversion [5]. Since then, the wavelength conversion based on population transfer has attracted extensive attention, and a series of conversion models have been developed, such as stimulated Raman adiabatic passage (STIRAP) [6-9] and Stark-chirped rapid adiabatic passage (SCRAP) [10, 11].

  In atomic system, SCRAP uses Stark shift caused by Stark field to produce energy level crossing, which realizes effective population transition from initial state to target state [6, 12]. Compared with STIRAP, it does not need two-photon resonance, and can be applied to multiphoton transmission with minimum intermediate state [7, 12]. Inspired by the application of STIRAP in QPM wavelength conversion [13], our group proposed a cascaded TWM process model, in which the conversion wavelength can cover the transparent range of nonlinear media [14, 15]. It also shows robustness to wavelength conversion conditions. It should be noted that the wavelength conversion of scrap mainly depends on the energy level crossing generated by nonlinear crystal coupling modulation [16]. However, the wavelength conversion scheme has strict requirements on the coupling coefficient. The corresponding crystal structure of this scheme is also limited, which is not conducive to the realization.

  In this study, we apply the two-peak SCRAP method to nonlinear two-cascaded wavelength conversion and focus on the influence of the coupling coefficient on the conversion process. First, we establish a cascaded wavelength conversion model based on two-peak SCRAP to realize the high-efficiency conversion process from a near-infrared laser to a mid-infrared laser. The influence of the modulation parameter on the two TWM cascaded wavelength conversion process is explored. Second, we simulate and discuss the effect of the intuitive coupling order and counterintuitive coupling order on the conversion efficiency. Finally, we discuss the wide bandwidth conversion feature of two-peak SCRAP cascaded wavelength conversion and compare the feature with wavelength conversion process based on STIRAP theory. We show that the two-peak SCRAP cascaded wavelength conversion process has the advantage of wide broadband conversion. The details are as follows.

## II. CASCADED WAVELENGTH CONVERSION MODEL BASED ON THE TWO-PEAK SCRAP THEORY

The Λ-type SCRAP population conversion process has many similarities with the cascaded TWM process. Three laser pulses are needed in SCRAP theory to achieve complete population transfer among three states in atomic system, which is shown in Figure 1. The pump pulse excites the particles from state $|1\rangle$ to state $|2\rangle$, and the Stokes pulse drives the particles from state $|2\rangle$ to state $|3\rangle$. During the transition, one steady off-resonant Stark pulse facilitates adiabatic sweep via a Stark shift, creating level crossings among the three states [11, 17]. In the cascaded TWM process, intermediate frequency $\omega_2$ is generated under the pump frequency $\omega_{p1}$ and input frequency $\omega_1$ after the first TWM process, then the intermediate frequency takes part in the second TWM process with the pump frequency $\omega_{p2}$ to obtain the output frequency $\omega_3$.

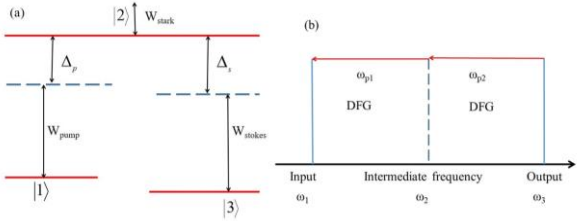

**FIGURE 1.** (a) Λ-type three-photon excitation level system. (b) Schematic diagram of two-cascaded DFG processes.

The Hamiltonian of the laser-excited three-level quantum system is expressed as follows [7]:

$$H(t) = \hbar \begin{bmatrix} 0 & \Omega_p(t) & 0 \\ \Omega_p(t) & \Delta_p + S_2(t) & \Omega_s(t) \\ 0 & \Omega_s(t) & \Delta_p - \Delta_s \end{bmatrix} \quad (1)$$

Where $\Omega_p(t)$ and $\Omega_s(t)$ are Rabi frequencies associated with the pump and Stokes fields, respectively. $S_2(t)$ represents for the Stark shift caused by Stark pulse. The parameters $\Delta_p$ and $\Delta_s$ are detuning induced by pump laser and Stokes laser.

In the undepleted pump approximation, there is dynamic symmetry in the three-state quantum system and two-cascaded TWM process[5]. Three states in quantum system corresponds to three wavelength, respectively. Time t is equal to the propagation length z. Rabbi frequencies are equivalent to the coupling parameters. The phase mismatches reflect the photon detunings. In SCRAP process, $S_2(t)$ produces the adiabatic crossings of the energy levels to achieve the population transfer from the initial state to the final state. In cascaded TWM process, the parameter $\kappa_c(z)$ connects the two TWM processes. Table 1 shows the approximate parameters in detail：

TABLE I.
ANALOGY BETWEEN THE CASCADED TWM PROCESSES IN THE UNDEPLETED PUMP APPROXIMATION AND THE DYNAMICS OF A TWO LEVEL ATOMIC SYSTEM, INDUCED BY COHERENT LIGHT[18].

| Parameter | Rapid adiabatic passage | TWM process |
|---|---|---|
| Evolution parameter | Time | Z axis |
| Ground/excited state population | $|a_g|^2, |a_e|^2$ | $|\varphi_1|^2, |\varphi_3|^2$ |
| Energy difference | $\omega_0 = \omega_{fg}$ | $n(\omega_2)\omega_2/c$ |
| Detuning/phase mismatch | $\Delta$ | $\Delta k$ |
| 'Rabi' frequency | $\Omega_0 = \mu E_{in}/\hbar$ | $\kappa = 4\pi\omega_1\omega_3\chi^{(2)}E_2/\sqrt{k_1k_3}c^2$ |

Based on the analogy and analysis of the two cascaded TWM process and the SCRAP process, we can describe the two-peak SCRAP cascaded wavelength conversion process as follows [3, 10, 18]:

$$\frac{d}{dz}\begin{bmatrix} \varphi_1 \\ \varphi_2 \\ \varphi_3 \end{bmatrix} = i \begin{bmatrix} 0 & \kappa_{12}(z)e^{-i\Delta k_1 z} & 0 \\ \kappa_{21}(z)e^{i\Delta k_1 z} & \Delta k_2 + \kappa_c(z)e^{-i\Delta k_3 z} & \kappa_{23}(z)e^{-i\Delta k_2 z} \\ 0 & \kappa_{32}(z)e^{i\Delta k_2 z} & \Delta k_3 \end{bmatrix}\begin{bmatrix} \varphi_1 \\ \varphi_2 \\ \varphi_3 \end{bmatrix} \quad (2)$$

An undepleted pump approximation is adopted in this process. $\varphi_j$ is the wave's complex amplitude with frequency $\omega_j$ (j=1,2,3), $\kappa_c(z)$ is the coupling-coefficient modulation parameter in the two-cascaded TWM wavelength conversion processes, and $\Delta k_j$ (j=1,2) represents the phase mismatch between the two-cascaded TWM conversion processes, where $\Delta k_1 = k_1 \pm k_{p1} - k_2$ and $\Delta k_2 = k_2 \pm k_{p2} - k_3$. The top sign $\Delta k_j$ corresponds to the sum-frequency generation (SFG) process, and the other corresponds to the DFG process. $\Delta k_3$ is the total phase mismatch between the two-cascaded wavelength conversion process, where $\Delta k_3 = \Delta k_1 + \Delta k_2$.

The properties of the two-cascaded wavelength conversion process can be discussed by considering (2), where $\kappa_{ij}$ is the sufficient coupling coefficient strength between $\omega_i$ and $\omega_j$, which is defined by $\kappa_{12} = \left[\chi^{(2)}(\omega_1, \omega_{p1}; \omega_2)\omega_1^2/k_1 c^2\right]\left[\text{Re}\{A_{p1}\} \mp i \text{Im}\{A_{p1}\}\right]$. In this part, $\omega_1$, $\omega_2$, and $\omega_{p1}$ are the frequencies of the input wavelength, the intermediate wavelength, and the pump wavelength during the first TWM process, respectively; $\kappa_{23} = \left[\chi^{(2)}(\omega_2, \omega_{p2}; \omega_3)\omega_2^2/k_2 c^2\right]\left[\text{Re}\{A_{p2}\} \mp i \text{Im}\{A_{p2}\}\right]$, where $\omega_3$ is the frequency of the output wavelength, and $\omega_{p2}$ is the frequency of the other pump wavelength. In the equations, $A_{p1}$ and $A_{p2}$ are the complex amplitude envelopes of the pump fields' amplitudes, $\chi^{(2)}$ is the second-order nonlinear coefficient of the material. Both the equations of $\kappa_{ij}$ satisfy the condition of $\kappa_{ij} = \left(\omega_i^2 k_j / \omega_j^2 k_i\right)k_{ji}^*$.

In the process of the cascaded TWM process, $\kappa_c(z)$, $\kappa_{12}(z)$, and $\kappa_{23}(z)$ are designed as follows:

$$\begin{aligned}
\kappa_{12}(z) &= \kappa_{12} e^{-(z+s_1)^2/d_1^2} \\
\kappa_{23}(z) &= \kappa_{23} e^{-(z+s_2)^2/d_2^2} \\
\kappa_c(z) &= \frac{1}{2}\kappa_c \left( e^{-(z+s_c/2)^2/d_3^2} + e^{-(z-s_c/2)^2/d_4^2} \right) \\
\kappa_c &= (\kappa_{12}\kappa_{21} + \kappa_{23}\kappa_{32})^{1/2}
\end{aligned} \quad (3)$$

In function (3), $s_j$ and $d_j$ are the coupling delay and width parameters, and $S_c/2$ and $-S_c/2$ are the delays of the two peaks of the parameter $\kappa_c(z)/\kappa_c$ along the propagation length. Parameter $\kappa_c(z)$ connects the two coupling parameters modulating the two cascaded processes. $S_c$ determines the distance between the two peaks, which is mainly studied in this paper. The value of $S_c$ determines the shape of coupling coefficient $\kappa_c(z)$ in the propagation length. When $S_c = 0$, the two peaks of $\kappa_c(z)/\kappa_c$ coincide, and the coupling coefficient shows a typical single peak Gaussian shape. $s_j$ determines the coupling order; when $s_1 < s_2$, the coupling process is in counterintuitive order; when $s_1 > s_2$, it is in intuitive order.

Under the adiabatic conditions of SCRAP theory, when the wavelength of the input and pump laser is appropriately selected, the conversion process meets the phase-matching condition, $\Delta k_1 = \Delta k_2 = \Delta k_3 = 0$. In this instance, the corresponding normalized eigenvectors can be described as [19]:

$$|g_0\rangle = \frac{1}{\kappa_s}\begin{bmatrix} \kappa_{32} \\ 0 \\ -\kappa_{12} \end{bmatrix}, \quad |g_\pm\rangle = \frac{1}{\sqrt{\kappa_s^2 + \left(\frac{1\pm\sqrt{5}}{2}\right)^2 \kappa_c^2}} \begin{bmatrix} \kappa_{12} \\ \frac{1\pm\sqrt{5}}{2}\kappa_c \\ \kappa_{32} \end{bmatrix} \quad (4)$$

In the function, the parameters are defined as $\kappa_s = \sqrt{\kappa_{12}^2 + \kappa_{32}^2}$. The mixing angle is described as $\theta = \tan^{-1}(\kappa_{12}/\kappa_{32})$. Thus, the adiabatic state $|g_0\rangle$ can be further described as follows:

$$|g_0\rangle = \begin{bmatrix} \cos\theta \\ 0 \\ -\sin\theta \end{bmatrix} \quad (5)$$

The coupling between the target eigenstate and the other eigenstates is much smaller than the difference among them to keep the process in the state of adiabatic evolution. The adiabatic condition can be described as follows [19]:

$$\left|\frac{d\theta}{dz}\right| << \frac{\kappa_c}{\kappa_s\sqrt{\kappa_s^2 + \left(\frac{1\pm\sqrt{5}}{2}\right)^2 \kappa_c^2}} \quad (6)$$

Phase reversal quasi-phase-matching (PRQPM) is used to

### III. Numerical simulation results and discussion

We demonstrate a two TWM cascaded process based on the two-peak SCRAP scheme by performing numerical simulations in this study. The TWM processes are assumed to adopt an undepleted pump approximation so that the associated coupling function is able to be modulated as modulate the second-order nonlinear coefficient $\chi^{(2)}$ in the optic crystals [20-22]. Pragati *et al.* designed a series of structures in the optical crystal by mainly using two different quasi-phase-matching (QPM) grating structures in a hybrid planar-channel waveguide to achieve the two cascaded wavelength conversion process with STIRAP mechanism effects, and they realized the effect of two coupling coefficients in the construction of an adiabatic process in the action length[23]. The arrangement of the periodic inversion domain structure in the action distance affects the coupling coefficient in the process of wavelength conversion. Therefore, changing the array rule of the domain structure modulate the coupling coefficient. In the crystal structure we apply, we use three kinds of staggered grating structures. To achieve the effect of the target coupling coefficient, the spatial distribution of the grating structure is adjusted by using PRQPM technology. In this study, we use the connection coupling coefficient of two-peak Gaussian wave shape. Figure 2 shows the optical crystal structure we designed in this study. The gray region in the figure represents the conventional structure region, and the different colored regions correspond to domain inversion structure regions with different functions. The orange region with a long propagation distance distribution realizes coupling coefficient $\kappa_c(z)$, while the green region and the blue region with a short propagation distance represent coupling coefficients $\kappa_{12}(z)$ and $\kappa_{23}(z)$. Coefficients $s_j$ and $d_j$ can be adapted by the translation and amplification of the domain inversion regions in the crystal. The chirped domain inversion structure can realize the spatial distribution of the coupling coefficient in the process of cascaded wavelength conversion, because the relative position and distribution length of the inversion structure are related to the parameters of the coupling coefficient. Therefore, we can design the inversion domain region according to the spatial distribution of different coupling coefficients.

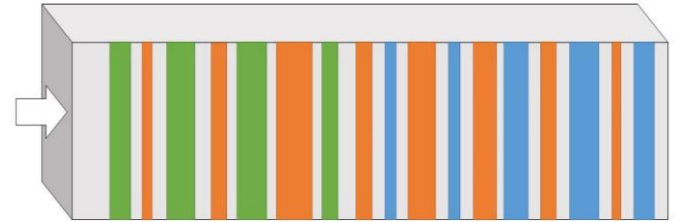

**FIGURE 2.** The light-colored and colored parts show the positive and negative domains respectively. Different color parts represent different QPM gratings for wavelength conversion processes.

expected.

We use two DFG cascaded processes to demonstrate the wavelength conversion process. In the whole process, signal wavelength $\lambda_1$ is 1064 nm; pump wavelengths $\lambda_{p1}$ and $\lambda_{p2}$ are 2700nm and 3000 nm, respectively. Intermediate wavelength $\lambda_2$ is 1750 nm, and output wavelength $\lambda_3$ is 4200 nm. The sequence of the wavelength conversion

process is adopted as follows:
$$\lambda_1(1,064nm) - \lambda_{p1}(2,700nm) \to \lambda_2(1,750nm)$$
$$\lambda_2(1,750nm) - \lambda_{p2}(3,000nm) \to \lambda_3(4,200nm)$$

In the simulation, we choose the periodic polarization LiNbO$_3$ (PPLN) crystal as the nonlinear medium. The lithium niobate crystal has excellent nonlinear optical properties and is a basic functional material used in various fields of optoelectronics [24]. Other optical crystal grating structures can also realize the two-peak cascaded wavelength conversion process by choosing appropriate wavelengths and coupling coefficients. In this paper, the two-peak SCRAP cascaded wavelength conversion optical design model is demonstrated by using the PPLN material as an example. The crystal's nonlinear coefficient is 28pm/V, and the crystal length is 40 mm. The refractive indices in the relevant coupling parameters are calculated from the Sellmeier equation at a temperature of 373 K [24]. The input pulse intensity is 100 MW/cm$^2$, and the pump pulses have peak intensities of 100 GW/cm$^2$ and 800 GW/cm$^2$. Modulation parameter $S_c$ is set as 16 mm.

According to the two-cascaded wavelength conversion model that we established, as shown in (2) and (3), we simulated the two TWM cascaded conversion process along the nonlinear crystal, as shown in Figure 3.

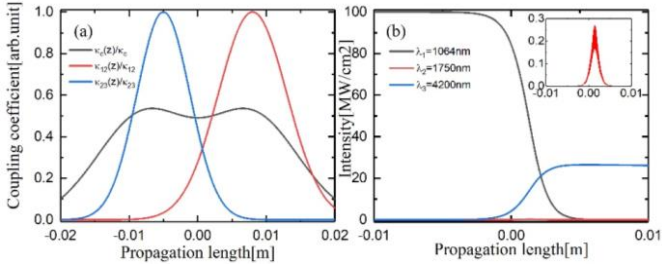

**FIGURE 3.** (a) Coupling-coefficient modulation along the propagation length. The zero point on the horizontal axis represents the center of the process in the crystal. The relative position between the coupling coefficients required for the TWM cascaded conversion process is shown. (b) Three-wavelength intensity variations along with the nonlinear crystal. The enlarged graph of the intermediate wavelength intensity is located in the whole graph's upper right part.

In Figure 3, the coordinate origin of the propagation length is set in the middle of the crystal length, the area with negative coordinates represents the former part of the crystal, and the area with positive coordinates represents the latter part of the crystal. In the figures, the light pulse is injected into the crystal at -0.02 m, the multiwave mixing process occurs in the three period domain inversion structure, and the crystal is emitted at 0.02 m. Figure 3. (a) shows the modulation parameter under counterintuitive coupling order along the propagation length. In the process of TWM, $\kappa_{12}(z)/\kappa_{12}$ and $\kappa_{23}(z)/\kappa_{23}$ reach their maximum values after and before the center of $\kappa_c(z)/\kappa_c$ respectively and cross each other at the midpoint of the crystal length. Through the two-peak modulation process, the two TMW processes are connected to achieve highly efficient wavelength conversion.

We simulate the cascaded wavelength conversion processes along the nonlinear crystal, as shown in Figure 3 (b). Here, the two TWM cascaded conversion process starts at -5mm and end at 4mm along the crystal direction. They are affected by the coupling delay and width coefficients of the coupling parameters. This process is consistent with our description of energy transfer and conversion in (2). The inset in Figure 3 (b) shows the variation in the intensity at the intermediate wavelength. The inset shows that the intermediate wavelength intensity is 0.27% of the input intensity, which means there is no significant intermediate-wavelength generation. The input wavelength produces complete conversion, and the output wavelength intensity is 25.3 MW/cm$^2$, which satisfies the proportional relationship $I_3/I_1 = \lambda_1/\lambda_3$ required for complete conversion between different wavelengths. The above simulation results are obtained when the system adopts fixed and appropriate parameters. The adiabatic crossing of appropriate coupling modulation parameters is a necessary condition to ensure the efficiency of wavelength conversion [5, 7, 18, 25]. Figure 3 (a) shows the condition of a counterintuitive coupling order of SCRAP mechanism. Under this circumstance, the coupling between the intermediate wavelength $\lambda_2$ and output wavelength $\lambda_3$ precedes the coupling between input wavelength $\lambda_1$ and intermediate wavelength $\lambda_2$, thus the energy can be converted from the signal to the output, while the intermediate wavelength $\lambda_2$ remains almost zero [7].

Next, we will discuss the influence of the separation coupling coefficient on the conversion results. We change the modulation parameter $S_c$ from 0 m to 0.02 m to obtain coupling parameter $\kappa_c(z)$ and explore the influence on the cascaded wavelength conversion. Parameter $S_c$ can be changed by adapting the arrangement of the corresponding inversion domain structure in the propagation length. Figure 4. (a) shows the evolution of the intermediate wavelength intensity along the propagation length when modulation parameter $S_c$ changes in the range we set. The intermediate wavelength intensity is inversely proportional to the increase in $S_c$ from -2 mm to 3 mm along the propagation length. Figure 4 (b) shows the specific differences in wavelength intensity, and the range of parameters selected is consistent with Figure 4 (a). The values of each vertical column shown in the figure are the difference in the intensities at the same distance position. When $s_c = 0$, coupling modulation coefficient $\kappa_c(z)$ is in the one-peak Gaussian state. When modulation parameter $S_c$ increases slightly from 0 m, the intensity distribution of the intermediate wavelength has little change. As the value of $S_c$ continues to increase, the intermediate wavelength intensity first decreases slightly at the peak area, compared with the content shown in Figure 4 (a) on the left, while the other positions still maintain the same wavelength intensity. Then $S_c$ increases to a larger value, and the effect of two peaks' separation on the conversion process becomes more obvious. The maximum value of the intensity difference increases with the value of $S_c$, and the region of difference on intensity also expands.

Within this range we set, the increase in the $S_c$ value steadily reduce the intensity of the intermediate wavelength, and the difference area is mainly near the peak value. Next, we will discuss the influence of the coupling coefficient $S_c$ on the conversion efficiency in the process of wavelength conversion.

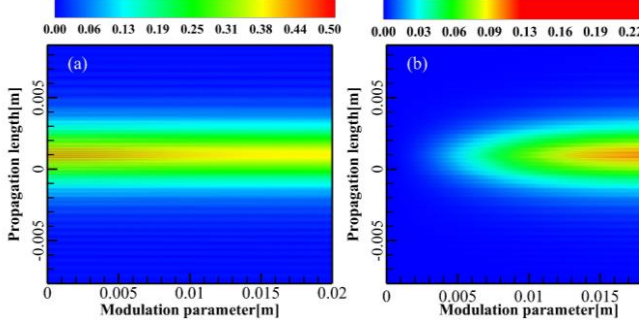

FIGURE 4. (a) The intermediate wavelength intensity of the two-cascaded wavelength conversion process along the propagation length with respect to modulation parameter $S_c$. The color represents the value of the intermediate wavelength intensity. (b) The difference between the intermediate light intensity vary with the tuning parameter and the propagation length.

We explore the influence of the modulation coefficient on the spatial variation of the conversion efficiency. The results are shown in Figure 5. First, when other coupling parameters remain unchanged, separation modulation parameter $S_c$ increases gradually from 0 m to 0.02 m, and the conversion efficiency does not change with the increase in coupling parameters. This analysis result is the same as that discussed in (4) and (5). When coupling parameters $\kappa_{12}(z)$ and $\kappa_{23}(z)$ are constant, the adiabatic conversion result is independent of the influence of $\kappa_c(z)/\kappa_c$, the conversion efficiency is not affected by the modulation parameter $S_c$. Then, we take modulation parameter $S_c$ as 0.01 m and change coupling parameters $S_1$ and $S_2$, so that the process completes the intuitive coupling sequence and gradually changes to the counterintuitive coupling sequence under the condition of constant $S_c$. The concrete results are shown in Figure 5 (b). Take the diagonal line from the graph's bottom left to the graph's upper right corner as the dividing line. In the upper left area, $S_1$ is larger than $S_2$, and the coupling coefficient is in counterintuitive coupling order; in the lower right area, $S_1$ is less than $S_2$, and the coupling coefficient is in intuitive coupling order. Cascaded wavelength conversion under two-peak SCRAP can achieve a high conversion efficiency in a wide range of intuitive sequences and counterintuitive sequences. The figure's dividing line represents the condition where the distribution of the coupling modulation coefficients and coincides with the propagation length. In equation (6), we discussed the adiabatic condition. On the one hand, the difference between the eigenstates has a large ratio relative to the change in mixing angle $\theta$ in most sequences. On the other hand, in the region of low efficiency rates, the conversion process cannot achieve the adiabatic condition. In the region of homologous coupling coefficients, the conversion efficiency will be relatively low, but there are some specific points can form a 'channel' between the adjacent two coupling sequence regions. Although these special cases have high conversion rates, they do not satisfy the adiabatic condition. Therefore, the special cases are unstable and is not under adiabatic passage scheme.

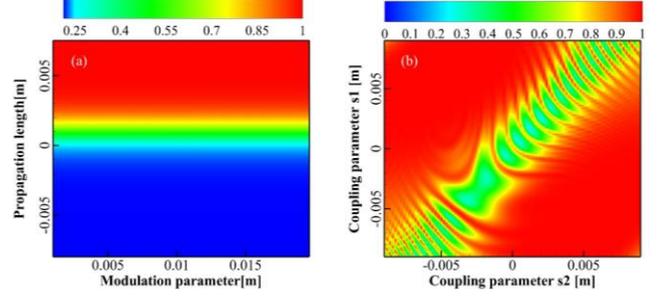

FIGURE 5. (a) The efficiency of two-cascaded wavelength conversion process along the propagation length under different modulation parameters. The color change along the propagation length corresponds to the conversion efficiency of the corresponding position. (b) The efficiency of two-cascaded wavelength conversion process at propagation length 19 mm for different coupling parameters $S_1$ and $S_2$; at the same time, the modulation parameter is 10mm.

We compared the cascaded wavelength conversion method based on STIRAP with the same wavelength and obtained the conversion efficiency results. In the case under STIRAP scheme, the PPLN crystal consists of two domain structures. We set different coupling delay coefficients and width coefficients and combine their spatial distribution. The conversion results are shown in Figure 6. The results show that in the wavelength conversion scheme demonstrated in this paper, the two-cascaded wavelength conversion efficiency under the STIRAP mechanism is less than 10%. The low conversion rate is due to the high degree of phase mismatching of wavelength combination, and it is difficult to achieve efficient wavelength conversion with a double domain inversion structure without coupling parameter $\kappa_c(z)$. When the incident wavelength changes in a certain range, the wavelength conversion model based on the two-peak SCRAP scheme has good adaptability for the processing of wide bandwidth input wavelength. Figure 6 (b) shows that when other parameters remain unchanged in the cascaded wavelength conversion process, the input wavelength in the range of 1046 nm to 1077 nm has a relatively ideal conversion effect, and the conversion efficiency is above 95%. When the input wavelength is between 1051 nm and 1069 nm, the theoretical conversion rate is above 99%. When the input wavelength is beyond this range, the conversion efficiency will decrease rapidly with increasing distance from the efficient conversion wavelength range. In the actual crystal structure, the multidomain structure will reduce the intensity of the target wavelength. However, the wavelength conversion scheme based on the two-peak SCRAP mechanism can be completed in a short length, so the dissipation caused by the crystal structure can be reduced. In general, the scheme based on the two-peak SCRAP scheme has good wide bandwidth conversion ability, which can handle large wavelength detuning and wide input wavelength bandwidth.

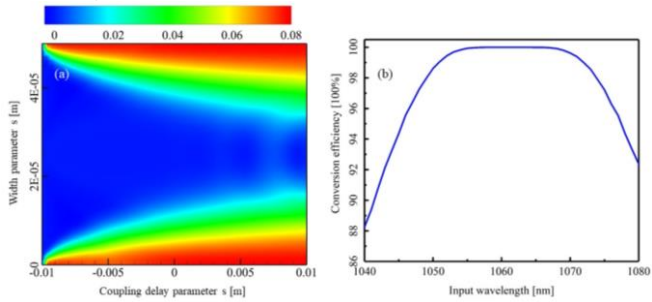

FIGURE 6. (a) Under the same phase-mismatch condition, the wavelength conversion based on the STIRAP mechanism. (b) At different input wavelengths, the wavelength conversion rate based on the two-peak SCRAP mechanism.

## IV. CONCLUSION

In this paper, we established a numerical model of two TWM cascaded wavelength conversion process based on the two-peak SCRAP scheme. We adopt a hybrid periodic domain inversion structure to realize the two cascaded wavelength conversion process from 1064nm to 4200nm. In the simulated model, the incident wavelength conversion rate is 100%, and the maximum intermediate wavelength intensity is 0.27% of the incident wavelength intensity. The conversion result is effective and robust. We simulate and study the effect of different coupling modulation coefficients on the conversion results. The coupling modulation is modulated by adjusting the spatial position of the inversion domain structure in the optical crystal. We show that this cascaded wavelength conversion scheme has good wide bandwidth conversion ability, and the conversion effect is better than the wavelength conversion scheme based on the STIRAP mechanism.

The numerical analysis of the two-peak SCRAP cascaded wavelength conversion process shows that when the two peaks' position of coupling coefficient $S_c$ increase from 0 mm to 20 mm, the maximum intensity of the intermediate wavelength decreases, which effectively reduces the loss of the intermediate wavelength. When the input wavelength is in the range of 1046nm ~ 1077nm, the efficiency of cascaded wavelength conversion process is more than 95%. The cascaded wavelength conversion efficiency is not affected by the distance of the peak spacing. It achieves 100% conversion efficiency in intuitive coupling sequences and counterintuitive coupling sequences, which means that the inversion domain structure in the crystal can flexibly combine the spatial position and has good transmission efficiency. The three periodic inversion domain structure and two-peak SCRAP scheme two-cascaded wavelength conversion technology can reduce the light intensity loss in the process, have a good wide bandwidth conversion rate, adapt to high phase mismatch conditions, and reduce the requirements of crystal structure.

We expect this work could provide a theoretical basis for photonic crystal structure and efficient wavelength generation method of laser sources.

**Acknowledgment**

This work is supported by the Guangdong Provincial Natural Science Foundation of China (Grant Nos.2015A030313383).